\documentclass[10pt]{article}
\usepackage{times}

\usepackage[papersize={8.5in,11in}, 
            total={5.6in,8.0in}, 
            centering]{geometry}

\usepackage{amsmath,amssymb,amsfonts}

\usepackage{braket}
\usepackage{parskip}
\usepackage{comment}
\usepackage{multirow}
\begin{document}

\title{Pattern Based Stabilizer Tripware}

\author{Mehedi Hasan Rumi \\
Dept. of Physics, University of Dhaka \\
Orcid id: 0009-0007-6234-6049}

\date{}

\maketitle

\begin{abstract}

This paper introduces a quantum-classical hybrid key distribution protocol applicable to any non-CSS stabilizer-based quantum error-correcting code. Named the \textit{Pattern-Based Stabilizer Tripwire}, the framework embeds logical data directly within the commutation relationships of physical Pauli operators while masking the underlying qubit layout using a dynamic, secret automorphism group of physical permutations. This coordinate-frame alignment ensures a 100\% sifting efficiency during normal operation without discarding transmission slots. However, any active interception by an eavesdropper introduces a destructive structural logical error. Because the adversary cannot align with the hidden spatial frame transformation, the receiver’s look-up decoder is blinded by a mismatched syndrome. This architectural misalignment forces an uncorrectable 50\% Quantum Bit Error Rate upon interception, converting subtle anomalies into a deterministic cryptographic tripwire. To secure private key generation, the algorithm utilizes a piece-wise conditional XOR framework to dynamically modulate data commutation relationships via public classical error syndromes. By executing an algebraic XOR operation that selectively fuses or withholds local channel noise syndromes and targeted, commutation-changing operator sub-vectors, the public classical channel functions as an information-theoretic blinding mask.
\end{abstract}

\section{Introduction}

Quantum Key Distribution [1] protocols traditionally rely on non-orthogonal physical states to detect eavesdroppers. Due to quantum uncertainty, a significant amount of raw bits must be discarded during the sifting phase. While this process ensures the privacy of the communication, the resulting Quantum Bit Error Rate (QBER) remains low enough that a sophisticated eavesdropper might still blend into a high-noise environmental channel. In this paper, I introduce a universal quantum algorithm applicable to any non-CSS stabilizer-based quantum error-correcting code within QKD systems. Because the algorithm follows deterministic, classical XOR-isomorphic logic, data can be transferred with 100\% sifting efficiency. Rather than focusing on hiding the data bits directly through state uncertainty, the protocol actively drives the QBER to its maximum theoretical limit of 50\% when an active eavesdropper is present, acting as a highly sensitive, structure-based intruder detection tripwire.

\section{Algebraic Framework}

Alice prepares a logical state $\lvert \psi \rangle_L$ corresponding to her bit $b_A \in \{0, 1\}$ within an $[[n, k, d]]$ non-CSS stabilizer code, where $n$ is the number of physical qubits, $k$ is the number of encoded logical qubits, and $d$ is the code distance. Rather than transmitting an uncorrupted codeword, she randomly selects a Pauli operator pattern $E_A$ that may commute ($[E_A, M] = 0$) or anti-commute ($\{E_A, M\} = 0$) with the logical measurement operator $M = Z_L$. The pattern $E_A$ must be an uncorrectable error having a physical weight greater than the code distance allows to fix, $w(E_A) > \lfloor (d-1)/2 \rfloor$. Within the code's Pauli space, the set of uncorrectable errors is chosen uniformly from a balanced pool evenly distributed between commuting and anti-commuting operators, ensuring an exact $50/50$ probability.

Alice restricts her choice of the initial Pauli operator $E_A$ to configurations that are as spatially asymmetric as possible. Fully symmetric operators (such as global tensor products $\bar{X}$ or $\bar{Z}$), which remain invariant under coordinate transformations, are strictly excluded from the selection pool. This ensures that any spatial permutation applied to the state forces a non-trivial transformation of the operator frame. Conversely, the chosen intra-block permutations $P_A, P_B \in \text{Aut}(S) \setminus \{I\}$ preserve the stabilizer generator group $S$ and leave the logical measurement observable $M = Z_L$ invariant up to stabilizer multiplication (for example, via a permutation-symmetric logical operator such as $Z_L = ZZZZZ$ in the $[[5,1,3]]$ code, or by restricting permutations to the logical-fixing subgroup of $\text{Aut}(S)$ in general non-CSS qLDPC codes). This structural invariance anchors the code space and logical readout basis during transmission, preventing basis collisions or uncontrolled quantum projections.

Before releasing the state into the channel, Alice uniformly selects a secret, non-trivial intra-block physical permutation $P_A \in \text{Aut}(S) \setminus \{I\}$ from the code's automorphism group (such as cyclic shifts or dihedral reflections in the $[[5,1,3]]$ code). Because $P_A$ belongs to $\text{Aut}(S)$, it maps the stabilizer group back onto itself while shifting the asymmetric error $E_A$ across physical coordinates to alter its syndrome.

Upon receiving the quantum state from Alice, Bob uniformly selects his own secret intra-block automorphism $P_B \in \text{Aut}(S) \setminus \{I\}$ and applies a random local Pauli operator $E_B$ to each logical block. Under the net intra-block permutation $P_{\text{net}} = P_B P_A$, the stabilizer generators in Bob's laboratory frame transform according to
$$S'_i = (P_B P_A) S_i (P_B P_A)^{-1} \in S,$$
keeping the stabilizer group intact. Accounting for any environmental channel noise $E_{\text{noise}}$ accumulated during transit, the composite operator $E_C = E_B \cdot P_B P_A(E_A)$ dictates the noise-free measurement outcome: when $[E_C, M] = 0$, Bob receives the true bit, and when $\{E_C, M\} = 0$, Bob receives the flipped bit.

\section{Key Generation Privacy}

The commutation and anti-commutation relationships of the Pauli operators $E_A$ and $E_B$ with respect to the logical measurement operator $M$ map isomorphically to a classical XOR operation. Let $c_A \in \{0, 1\}$ denote the parity bit indicating whether Alice's effective error operator commutes ($c_A = 0$) or anti-commutes ($c_A = 1$) with $M$, and let $c_B \in \{0, 1\}$ denote the corresponding commutation parity of Bob's local operator $E_B$. In a noise-free transmission, Bob's raw logical measurement outcome satisfies $b_B = b_A \oplus c_A \oplus c_B$. Consequently, Alice and Bob can seamlessly reconstruct their shared data post-measurement by communicating the commutation properties of their respective operators over the authenticated classical channel. Specifically, when Bob communicates the commutation relationship $c_B$ of $E_B$, Alice keeps her bit $b_A$ when the commutation relationship of $E_A$ matches that of $E_B$ ($c_A = c_B$), and flips her bit when they do not match ($c_A \neq c_B$).

However, because this parity tracking mirrors classical XOR logic, an eavesdropper passively listening to the classical channel could retroactively deduce information if the commutation relationship of $E_A$ were openly exposed. Deterministic algebra cannot simultaneously mask the data bit while publicly revealing its underlying commutation logic. While Alice and Bob intentionally disclose their raw bits during parameter estimation rounds to calculate the quantum bit error rate (QBER), key-generation rounds must conceal Alice's commutation parity completely from any passive or active observer.

To achieve unconditional privacy during key generation, Alice masks her commutation parity through Syndrome Fusion Masking, which relies on the scrambled error footprint $P_B P_A(E_A)$. After Bob broadcasts his chosen intra-block automorphism $P_B$, his isolated syndrome $s_B$, and his commutation parity $c_B$, Alice computes the expected syndrome $s'_B = s(P_B P_A(E_A))$ and isolates the discrepancy syndrome $s_{\text{noise}} = s_B \oplus s'_B$. Accounting for both natural environmental noise $s_{\text{env}}$ and any disturbance syndrome $s_{\text{Eve}}$ injected by an eavesdropper who mutates $E_A \mapsto E'_A$, this isolated syndrome decomposes as
$$s_{\text{noise}} = s_{\text{env}} \oplus s_{\text{Eve}}, \quad \text{where } s_{\text{Eve}} = s(P_B(E'_A)) \oplus s(P_B P_A(E_A)).$$
Whereas parameter estimation rounds always set $s_{\text{total}} = s_{\text{noise}}$, key-generation rounds allow Alice to construct the classical syndrome broadcast $s_{\text{total}}$ using two distinct ways:
\begin{enumerate}
    \item \textbf{Withholding Channel Noise ($C_1 = 0, C_2 = 1$):} Alice withholds $s_{\text{noise}}$ completely and broadcasts $s_{\text{total}} = s_{\text{part}}$, where $s_{\text{part}}$ represents the error syndrome of a correctable, low-weight portion $E_{\text{part}}$ extracted from or added to the uncorrectable error $E_A$ itself.
    \item \textbf{Transmitting or Combining Channel Noise ($C_1 = 1, C_2 \in \{0, 1\}$):} Alice includes $s_{\text{noise}}$ in her broadcast, either sending pure discrepancy syndrome ($s_{\text{total}} = s_{\text{noise}}$) or combining it with the partial error syndrome ($s_{\text{total}} = s_{\text{noise}} \oplus s_{\text{part}}$).
\end{enumerate}
Both modes are unified under the linear syndrome masking relation
$$s_{\text{total}} = C_1 \cdot s_{\text{noise}} \oplus C_2 \cdot s_{\text{part}},$$
where $C_1, C_2 \in \{0, 1\}$ are selected uniformly at random per round subject to $(C_1, C_2) \neq (0, 0)$.

This mechanism allows Alice to dynamically steer Bob's post-transmission key bit after the quantum communication phase has already finished. Because Alice already isolated $s_{\text{noise}} = s_B \oplus s'_B$ from Bob's broadcast, she knows the exact composite error operator $E_{\text{env}} \cdot E_{\text{Eve}}$ and its commutation parity with $M$. Rather than attempting to decode the high-weight uncorrectable error $E_A$ back to the all-zero syndrome, Bob applies a deterministic low-weight Pauli recovery operator $D(s_{\text{total}})$ corresponding strictly to $s_{\text{total}}$ before measuring $M$. When Alice sets $C_1 = 0$ and $C_2 = 1$, she leaves $s_{\text{noise}}$ uncompensated while directing Bob's decoder to apply $D(s_{\text{part}})$; when she sets $C_1 = 1$ and $C_2 = 1$, she folds $E_{\text{part}}$ into the recovery operation alongside $s_{\text{noise}}$. Whenever the net uncompensated error and partial recovery operator anti-commutes with $M$, Bob's measured logical bit flips dynamically, and Alice updates her private commutation parity $c_A$ and reconciled key bit $b'_A = b_A \oplus c_A \oplus c_B$ in unison so that both parties arrive at the exact same shared bit immediately.

For an eavesdropper observing the public classical broadcast $s_{\text{total}}$, the masking relation contains three independent unknown variables: the binary selector $C_1$, the binary selector $C_2$, and the underlying partial syndrome $s_{\text{part}}$ (alongside $s_{\text{noise}}$). Because a single syndrome vector $s_{\text{total}}$ leaves this linear system algebraically underdetermined, $s_{\text{part}}$ is strictly unsolvable for an eavesdropper. Furthermore, because $s_{\text{Eve}} \neq 0$ occurs with probability $1 - 2^{-(n-k)}$ whenever Eve disturbs $E_A$, an active attacker forces $s_{\text{noise}} \neq 0$ and cannot exploit the zero-noise $s_{\text{noise}} = 0$ regime; even on unattacked low-noise rounds where $s_{\text{noise}} = 0$ indicates $C_2 = 1$, $E_{\text{part}}$ is drawn equally from operators that commute and anti-commute with $M$, leaving Eve blind to whether Alice's commutation parity flipped. Consequently, holding an entangled probe in a quantum memory yields zero retroactive advantage: Eve must interact with the qubits during transit before Bob collapses the stabilizers, and because $E_A$ is chosen uniformly per block while $s_{\text{total}}$ masks $s_{\text{part}}$ under the one-time selector $(C_1, C_2)$, Eve's post-transcript quantum memory states $\rho_E^{(0)}$ and $ \rho_E^{(1)}$ remain indistinguishable up to the syndrome collision bound $O(2^{-(n-k)})$.

\section{Intercept-and-Resend Security Analysis}

If an eavesdropper executes an intercept-and-resend attack during transmission, the original Pauli footprint $P_A(E_A)$ is permanently erased due to wavefunction collapse upon her projective measurement. Because Alice applies the secret intra-block automorphism $P_A \in \text{Aut}(S) \setminus \{I\}$, Eve's standard stabilizer measurements are misaligned relative to Alice's preparation frame and project the intercepted state onto a random syndrome outcome $s_{\text{eve}}$. To conceal her presence and restore the state to the all-zero syndrome vector before forwarding replacement qubits to Bob, Eve must guess and apply a corrective Pauli operator $E_E$ corresponding to $s_{\text{eve}}$. However, because the stabilizer syndrome map is highly degenerate across distinct Pauli errors and Eve does not know $P_A(E_A)$ or Bob's upcoming secret automorphism $P_B$, her injected operator $E_E$ diverges structurally from Alice's true footprint.

Once Bob receives the resent transmission and applies his secret intra-block automorphism $P_B \in \text{Aut}(S) \setminus \{I\}$, Eve's injected footprint transforms into $P_B(E_E)$, yielding an isolated syndrome $s_B = s(P_B(E_E))$ that fails to cancel with Alice's expected syndrome $s'_B = s(P_B P_A(E_A))$. This un-canceled structural mismatch generates the inequality
$$s_B \neq s'_B.$$
In an ideal, noise-free channel, this inequality acts as a deterministic cryptographic alarm during parameter estimation rounds, immediately exposing Eve's intrusion. Under realistic environmental channel noise, Alice interprets the discrepancy $s_B \oplus s'_B$ as natural channel fluctuations and broadcasts a reconciliation syndrome $s_{\text{total}}$ that directs Bob's deterministic decoder to apply an incompatible low-weight Pauli recovery operator $D(s_{\text{total}})$. Because this false recovery operator is applied to a state already mutated by $P_B(E_E)$, it fails to neutralize Eve's disturbance and instead combines with it to induce an uncorrectable logical bit-flip with probability $1/2$. This degenerate miscorrection cascade forces a catastrophic $50\%$ QBER across all intercepted rounds, effectively replacing Alice's transmission with random noise.

Theoretically, an eavesdropper can only evade syndrome detection within a single stabilizer block if her guessed operator $E_E$ happens to map under Bob's secret automorphism $P_B$ to the exact target syndrome $s'_B$. Because Eve cannot predict $P_B$ or $P_A(E_A)$ in advance, the probability of an accidental syndrome collision is bounded by $2^{-(n-k)}$ across uniform syndrome distributions. Crucially, the larger the available pool of distinct error syndromes ($2^{n-k}$), the lower the probability of Eve successfully matching the target syndrome, causing her probability of stealthy alignment to vanish asymptotically as a function of the code's parameters. Consequently, a single stabilizer block architecture is mathematically sufficient to expose an eavesdropper statistically over a sequence of transmissions. Moreover, even on the rare rounds where an accidental syndrome collision occurs, Syndrome Fusion Masking ($s_{\text{total}} = C_1 \cdot s_{\text{noise}} \oplus C_2 \cdot s_{\text{part}}$) keeps $s_{\text{part}}$ algebraically unsolvable, preventing Eve from determining whether Alice's commutation parity flipped and ensuring that Eve's mutual information $I(A:E)$ remains negligible.

\section{General Quantum Attacks}

Beyond classical intercept-and-resend strategies, a quantum adversary may avoid immediate projective measurement in transit and instead couple an ancilla quantum register to the flying qubits via a coherent entangling unitary $U_{\mathrm{Eve}}$, retaining the probe in a quantum memory until all classical reconciliation data ($P_B$, $s_B$, $c_B$, $s_{\mathrm{total}}$, and $P_{\mathrm{global}}$) are publicly broadcast. To understand how the protocol responds to arbitrary coherent interactions, consider how a physical unitary $U$ acts on a non-CSS stabilizer code carrying an existing uncorrectable Pauli footprint $E_A$. Extracting logical parity into an ancilla without disturbing the syndrome frame requires $U$ to commute with every stabilizer generator $S_i \in S$ while leaving the error syndrome invariant. Based on how $U$ couples the data block to the ancilla register, physical operations fall into three exhaustive categories:

\begin{enumerate}
    \item \textbf{Multi-Block Transversal Unitaries ($\mathcal{C}_1$, where $U = \bigotimes_{i=1}^n u_i$):} In CSS codes, stabilizer generators separate cleanly into pure $X$-type and $Z$-type operators, permitting transversal CNOT gates to extract logical parity without disturbing the code space. In non-CSS stabilizer codes, however, individual stabilizer generators contain mixed $X$, $Y$, and $Z$ Pauli operators acting on the same physical qubits. Applying a transversal two-qubit gate couples a stabilizer $S_i \otimes I_E$ to $S_i \otimes P_E$, where $P_E$ is a non-trivial Pauli string acting on the ancilla. Because the data qubits are entangled with the ancilla outside the stabilizer group $S$, subsequent measurement or partial tracing of the ancilla projects the data block into a distorted error operator $E'_A \neq E_A$, shifting the syndrome frame and exposing the interaction.

    \item \textbf{General Non-Transversal Unitaries ($\mathcal{C}_2 = \mathcal{U}_{\mathrm{total}} \setminus (\mathcal{C}_1 \cup \mathcal{C}_3)$):} Defined as the exact set-theoretic complement of Categories $1$ and $3$, this class encompasses every remaining multi-qubit entangling unitary in $\mathcal{U}_{\mathrm{total}}$, including gates that act partially on the stabilizer space and partially on the logical space. Whenever a unitary in $\mathcal{C}_2$ fails to commute with the stabilizer normalizer $N(S)$, its action mutates the existing error footprint $P_A(E_A)$ into a different Pauli string or a coherent superposition of altered error operators, shifting its physical weight and stabilizer syndrome ($s(E'_A) \neq s(P_A(E_A))$).

    \item \textbf{$n$-to-$1$ Parity Unitaries ($\mathcal{C}_3$):} The sole unitary class capable of preserving the stabilizer syndrome frame under ideal conditions is a controlled logical phase-kickback gate acting on an ancilla initialized in $\lvert + \rangle_E = \frac{1}{\sqrt{2}}(\lvert 0 \rangle_E + \lvert 1 \rangle_E)$,
    $$
    C(Z_L) = \lvert 0 \rangle\langle 0 \rvert_E \otimes I + \lvert 1 \rangle\langle 1 \rvert_E \otimes Z_L,
    $$
    (equivalently realized via multi-qubit CNOT parity extraction onto $\lvert 0 \rangle_E$), which couples the exact support of a single isolated logical operator $Z_L \in N(S) \setminus S$ to the ancilla qubit. Because $Z_L$ and $E_A$ are both $n$-qubit Pauli strings, they satisfy $Z_L E_A = (-1)^{c_A} E_A Z_L$ for $c_A \in \{0, 1\}$. Evolving the joint state and tracing out the ancilla yields
    $$
    \rho_{\mathrm{after}}^{(3)} = E_A \left( \frac{1}{2}\rho_L + \frac{1}{2} Z_L \rho_L Z_L \right) E_A^\dagger,
    $$
    which dephases the logical state into the measurement basis while leaving the physical error operator $E_A$ completely intact on the exterior.
\end{enumerate}

While Categories $1$ and $2$ inherently disturb the stabilizer syndrome frame, a permutation-symmetric logical operator such as $Z_L = ZZZZZ$ in a single-block $[[5, 1, 3]]$ code would ordinarily allow an adversary to execute a Category $3$ ($n$-to-$1$ parity) unitary $C(Z_L)$ in transit without altering $E_A$, and subsequently reconstruct the logical bit once the classical reconciliation transcript is broadcast. Neutralizing a Category $3$ attack is achieved by directly scrambling the logical operator support in transit so that no permutation-invariant logical subspace is exposed:

\begin{itemize}
    \item \textbf{Primary Multi-Logical-Qubit Architecture ($k \ge 2$ Non-CSS qLDPC Codes via $P_{\mathrm{global}}$):} Alice applies a secret inter-block physical qubit permutation $P_{\mathrm{global}}$ across a non-CSS qLDPC code encoding $k \ge 2$ logical qubits (or across a multi-block batch) prior to transmission, disclosing $P_{\mathrm{global}}$ only after the entire transmission arrives inside Bob's laboratory. Because $P_{\mathrm{global}}$ mixes the physical supports of distinct logical qubits, an eavesdropper lacking $P_{\mathrm{global}}$ in transit cannot isolate a single logical operator $Z_{L,i} \in N(S)$ of weight $w_L$. Her probability of blindly guessing a valid logical support scales inversely with the number of coordinate combinations, $\binom{n}{w_L}^{-1}$, and any misaligned parity gate she applies slices across partial supports of distinct logical operators while violating the permuted stabilizer generators ($\pi \notin \mathrm{Aut}(S)$).

    \item \textbf{Pedagogical Single-Logical-Qubit Illustration ($k = 1$ Dummy-Qubit Padding):} As an illustrative $k = 1$ toy model on a single $[[5, 1, 3]]$ block, Alice appends one unentangled dummy qubit prepared in the conjugate superposition state $\lvert + \rangle = \frac{1}{\sqrt{2}}(\lvert 0 \rangle + \lvert 1 \rangle)$ alongside the $n$ physical data qubits and randomly permutes all $n + 1$ positions, revealing the dummy-qubit coordinate only after Bob receives the qubits and measures the dummy wire in the $X$-basis. If Eve applies a blind $n$-qubit parity gate $C(Z_L)$ in transit, she aligns with the exact $n$ real data qubits with probability $1/(n + 1)$ (equal to $1/6$ for $[[5, 1, 3]]$). With probability $n/(n + 1)$ (equal to $5/6$), her parity gate misses a real data qubit and couples to the $\lvert + \rangle$ dummy qubit, randomizing her parity readout and collapsing the dummy wire into an $X$-basis error detected immediately by Bob without any additional basis-sifting overhead.
\end{itemize}

Because both $P_{\mathrm{global}}$ and dummy-qubit coordinate hiding eliminate any permutation-invariant logical subspace exposed in the channel, a coherent probe parameterized by an interaction angle $\theta \in [0, \pi/2]$,
$$
U_{\mathrm{Eve}}(\theta) = \cos\theta \, (I_{\mathrm{data}} \otimes I_E) - i\sin\theta \, (G_{\mathrm{data}} \otimes O_E),
$$
cannot extract logical parity without risking stabilizer disturbance with probability $p = \sin^2\theta$. Whenever $G_{\mathrm{data}}$ fails to commute with the true stabilizer frame, it shifts the error footprint in transit and injects a discrepancy $s_{\mathrm{Eve}} \neq 0$ into $s_{\mathrm{noise}} = s_{\mathrm{env}} \oplus s_{\mathrm{Eve}}$ with probability $1 - 2^{-(n-k)}$.

Finally, temporal causality prevents an eavesdropper holding a quantum memory from gaining a retroactive advantage after the classical transcript is broadcast. To correlate her ancilla with the transmitted state, Eve's probe $U_{\mathrm{Eve}}(\theta)$ must interact with the flying qubits during transit, strictly before Bob receives the batch, learns $P_{\mathrm{global}}$ (or the dummy-qubit coordinate), applies $P_B$, and performs projective stabilizer measurements. Once Bob measures the stabilizers, the state collapses and logs any syndrome discrepancy $s_B \neq s'_B$. Broadcasting the classical reconciliation data $(P_B, s_B, c_B, s_{\mathrm{total}}, P_{\mathrm{global}})$ afterwards provides no retroactive key recovery because $E_A$—along with its random commutation parity $c_A$—is never revealed, and Syndrome Fusion Masking ($s_{\mathrm{total}} = C_1 \cdot s_{\mathrm{noise}} \oplus C_2 \cdot s_{\mathrm{part}}$) conceals the steering parity under the one-time selector $(C_1, C_2)$ whenever a discrepancy $s_{\mathrm{noise}}$ is present.

Together, these mechanisms are divided across all coherent attack classes: non-CSS stabilizer entanglement combined with asymmetric $E_A$ Syndrome Fusion Masking neutralizes permutation-invariant transversal attacks ($\mathcal{C}_1$) which would otherwise commute with $P_{\mathrm{global}}$ as well as internal scrambling unitaries ($\mathcal{C}_2$), while delayed disclosure of $P_{\mathrm{global}}$ (or dummy-qubit coordinate hiding) eliminates the permutation-symmetric logical subspace required for stealth parity extraction ($\mathcal{C}_3$). Consequently, no unitary interaction in $\mathcal{U}_{\mathrm{total}}$ can extract logical key information without inducing a detectable syndrome disturbance prior to classical reconciliation.

\section{Conclusion}

To eliminate eavesdropper ambiguity, this paper introduces a quantum algorithm applicable to any non-CSS stabilizer-based quantum error-correcting code. In this algorithm, data can be transferred with 100\% sifting throughput. By forcing any coordinate misalignment to drive the localized QBER straight to 50\% of the qubits intercepted, the protocol ensures that an eavesdropper create high QBER signature above natural noise.

\end{document}